\documentclass[11pt]{article} 
\usepackage{hyperref} 
\pdfoutput=1 
 
\begin{document} 
 
\title{Three-dimensional pattern formation in suspensions of swimming micro-organisms} 
 
\author{Amir Alizadeh Pahlavan and David Saintillan \\ 
\\\vspace{6pt} Department of Mechanical Science and Engineering, \\ University of Illinois at Urbana-Champaign, Urbana IL 61801, USA} 
 
\maketitle 
 
 
\begin{abstract} 
Suspensions of self-propelled particles, such as swimming micro-organisms, are known to undergo complex dynamics as a result of hydrodynamic interactions. This fluid dynamics video (\href{http://ecommons.library.cornell.edu/bitstream/1813/14118/3/APS_DFD09_mpeg1.mov}{low-resolution}, \href{http://ecommons.library.cornell.edu/bitstream/1813/14118/2/APS_DFD09_mpeg2.mov}{high-resolution}) presents a numerical simulation of such a suspension, based on a kinetic theory recently developed by Saintillan and Shelley \cite{Saintillan08,Saintillan08b}. Starting from a nearly uniform and isotropic initial distribution, our simulations show the formation of strong density fluctuations, which merge and break up in time in a quasiperiodic fashion. These fluctuations are found to occur on the size of the simulation box, in agreement with a prediction from a linear stability analysis \cite{Saintillan08b}. In addition, the dynamics are characterized by complex chaotic flow fields, which result in efficient fluid mixing.

\end{abstract} 
 

\end{document}